\documentclass[aps,pra,reprint, showpacs, superscriptaddress]{revtex4-1}

\usepackage{epsfig,graphics,graphicx}
\usepackage{dcolumn}
\usepackage{bm}
\usepackage{amsmath,amssymb,amsthm}
\usepackage{centernot}
\usepackage{bbold}
\usepackage{soul}
\usepackage{mathtools}
\usepackage{fullpage}
\usepackage{units}
\usepackage[usenames,dvipsnames]{xcolor}
\usepackage{braket}
\usepackage{indentfirst}
\usepackage{lipsum}
\usepackage[export]{adjustbox}

\usepackage[utf8]{inputenc}
\usepackage[T1]{fontenc}

\usepackage{dsfont}

\usepackage{mathptmx}
\usepackage{amssymb}
\usepackage{amsmath}
\usepackage{latexsym}
\usepackage{amsfonts}

\usepackage{hyperref}
\hypersetup{pdfpagemode=UseNone}

\newtheorem{theorem}{Theorem}
\newtheorem{definition}[theorem]{Definition}

\newtheorem{lemma}[theorem]{Lemma}
\newtheorem{proposition}[theorem]{Proposition}

\newcounter{rem}
\newcommand{\mc}[1]{\mathcal{#1}}
\newcommand{\g}[1]{\mathfrak{#1}}

\DeclarePairedDelimiter\ceil{\lceil}{\rceil}

\def\>{\rangle}
\def\<{\langle}
\newcommand{\proj}[1]{| #1 \rangle\! \langle #1 |}
\newcommand{\borb}[2]{\left | #1 \rangle\!\langle #2 \right |}
\renewcommand{\rho}{\varrho}
\newcommand{\idty}{\mathds{1}}
\newcommand{\idtym}{\mathds{1}}
\def\tr{{\rm Tr}}

\def\ii{{\rm i}}

\def\textbf#1{{\bf #1}}
\newcommand{\Cx}{\mathbb{C}}

\newcommand{\Rl}{\mathds{R}}

\def\beq{\begin{equation}}
\def\eeq{\end{equation}}
\def\beqa{\begin{eqnarray}}
\def\eeqa{\end{eqnarray}}
\def\eea{\end{array}}
\def\bea{\begin{array}}
\newcommand{\bei}{\begin{itemize}}
\newcommand{\eei}{\end{itemize}}
\newcommand{\bee}{\begin{enumerate}}
\newcommand{\eee}{\end{enumerate}}
\def\bep{\begin{proposition}}
\def\eep{\end{proposition}}
\def\bel{\begin{lemma}}
\def\eel{\end{lemma}}
\def\bet{\begin{theorem}}
\def\eet{\end{theorem}}
\def\bed{\begin{definition}}
\def\eed{\end{definition}}
\newcommand{\fmelo}[1]{{\color{black} #1}}

\definecolor{cgreen}{RGB}{26, 199, 76}
\newcommand{\cris}[1]{{\color{black} #1}}

\usepackage{soul}
	
\begin{document}

\title{Emerging Dynamics Arising From Coarse-Grained Quantum 
Systems}

\author{Cristhiano Duarte}
\email{cristhiano@mat.ufmg.br}
\affiliation{Departamento de Matem\'{a}tica, Instituto de Ci\^{e}ncias Exatas, Universidade Federal de Minas Gerais, CP 702, CEP 30123-970, Belo Horizonte, Minas Gerais, Brazil.}

\author{Gabriel Dias Carvalho}
\email{gabrieldc@cbpf.br}
\affiliation{Centro Brasileiro de Pesquisas F\'isicas, Rio de Janeiro, Rio de Janeiro, CEP 22290-180}

\author{Nadja K. Bernardes}
\email{nadjakb@df.ufpe.br}
\affiliation{Departamento de F\'{\i}sica, Instituto	de Ci\^{e}ncias Exatas, Universidade Federal de Minas Gerais, CP 702, CEP 30123-970, Belo Horizonte, Minas Gerais, Brazil.}
\affiliation{\cris{Departamento de Física, Universidade Federal de 
Pernambuco, 50670-901 Recife, PE - Brazil.}}

\author{Fernando de Melo}
\email{fmelo@cbpf.br}
\affiliation{Centro Brasileiro de Pesquisas F\'isicas, Rio de Janeiro, Rio de Janeiro, CEP 22290-180}

\date{\today}


\begin{abstract}
\cris{The purpose of physics is} to describe nature from 
elementary 
particles all the way up to cosmological objects like cluster of galaxies 
and black holes. Although a unified description for all this 
spectrum of events is desirable, an one-theory-fits-all would be highly 
impractical. To not get lost in unnecessary details, effective descriptions 
are mandatory. Here we analyze what are the dynamics that may emerge 
from a full quantum description when one does not have access to all the 
degrees of freedom of a system. More concretely, we describe the properties 
of the dynamics that arise from quantum mechanics if one has 
only access to a coarse grained description of the system. We obtain that the effective 
maps are not necessarily of Kraus form, due to correlations between 
accessible and non-accessible degrees of freedom, and that the distance 
between two effective states may increase under the action of the effective 
map. We expect our framework to be useful for addressing 
questions such as the thermalization of closed quantum 
systems, as well as the description of measurements in quantum mechanics.
\end{abstract}

\pacs{03.65.Ta, 03.67.-a, 03.65.Yz}

\maketitle



\section{Introduction}

It is widely accepted that quantum mechanics provides currently the best desciption we have of the physical world.
However, the description of systems in our 
daily lives does not require the whole framework arising from 
quantum mechanics. In fact, our everyday life experiences heavily rely on 
effective 
(macroscopic) descriptions which are far less complex than their 
underlying intricate quantum characterization. 
For example, to  
describe the behavior of a macroscopic object, like the thermal  
expansions or compressions  of a rail line,  we do not need to specify the quantum states of all atoms 
composing such an object. 
In this situation we resort to the theory of 
thermodynamics~\cite{Callen}, 
which is probably the clearest example of effective theories. 
Although the systems treated within this theory are composed 
by many quantum interacting particles, macroscopic variables --such 
as temperature, volume, and pressure-- describe the systems well enough, 
allowing, for instance, for the design of thermal machines. 

The idea of different scales is central in physics. But, how  
does the description in one scale emerge from the description in a deeper 
scale?  Different ways of~\emph{coarse graining} the description of a system 
are often employed~\cite{Zwanzig, Mori, Castiglione, GellMann, 
Grabert} in order to ``zoom out'' from one level and obtain an 
effective description. Coarse grainings frequently appear in 
statistical physics~\cite{Grabert}, and are arguably the central tool in the 
renormalization method developed by Kadanoff and 
Wilson~\cite{Kadanoff,Wilson}. Nevertheless, some of these early methods are 
sometimes based on not so well controlled approximations or 
on projections, leading thus to ill-defined and/or probabilistic effective dynamics when applied to 
quantum systems.

\begin{figure}[t]
	\includegraphics[width=0.8\linewidth]{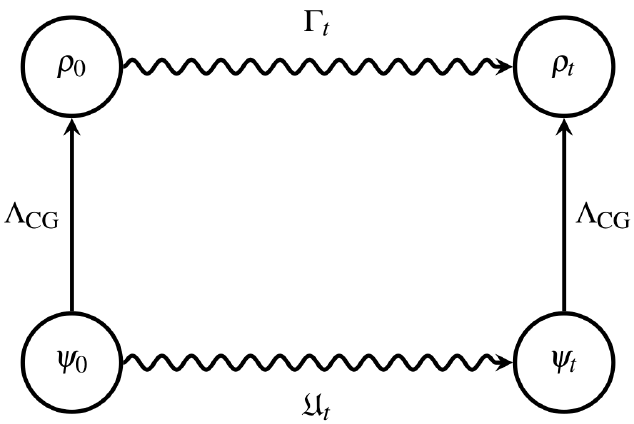}
	\caption{\footnotesize \textbf{Coarse graining induced dynamics.}
	Schematic diagram representing the different levels of description 
connected by a coarse graining.  Given an initial state of the 
system, \cris{with density operator} $\psi_0$, its evolution, 
$\g{U}_t$, and a coarse graining map 
$\Lambda_\text{CG}$, we want to determine what is the induced dynamics 
$\Gamma_t$, and its properties, such that $\Gamma_t \circ 
\Lambda_\text{CG}(\psi_0) = \Lambda_\text{CG} \circ \g{U}_t (\psi_0)$. 
}\label{FigureDiagramStates}
\end{figure}

In the last decades, with the birth of the quantum information field, 
various tools were developed to deal with many-body quantum
systems~\cite{nielsenchuang}. In particular, the theory of completely positive linear 
maps~\cite{WatrousLect,MarkWildeBook,MichaelGuide}, which 
aims at describing the most general transformations that can be applied to a 
system (including the most general time evolution), became well established. 
This has been accompanied and supported by the 
formalization and development of a theory for quantum 
correlations~\cite{horodecki09}, and by efficient descriptions of 
many-body quantum states~\cite{MPS, Orus}. The goal of the present 
contribution is  to employ some of these tools in order to obtain 
effective descriptions of quantum systems and their dynamics. More 
concretely, see Fig.~\ref{FigureDiagramStates}, given a system in the
state \cris{represented by a density operator} $\psi_0$ evolving by the 
unitary map $\g{U}_t$; what is the dynamics 
$\Gamma_t$ induced by a coarse graining $\Lambda_\text{CG}$? 
What types of dynamics might emerge when we departure from a 
full quantum description of the systems? 

In what follows we present a framework to address these questions. Its 
construction is closely related to that of  open quantum 
systems~\cite{breuer, Carteret08, Buzek01, RivasHuelgaPlenio}. In fact, concepts like the 
correlation between system and environment, and maps divisibility will 
play an important role here as well. Nevertheless, our framework encompasses 
and generalizes this previous formalism, as ours can be used in many other 
situations. It can, for instance, be used to describe closed systems from 
which just partial information is available, what might play a significant 
role in the thermalization of closed quantum 
systems~\cite{Eisert15, FaistThesis}. Our work is also 
related to recent articles by Kofler and Brukner~\cite{kofler1, kofler2}. In 
these articles the authors analyze the effect of coarse-grained measurements 
in order to explain the emergence of the classical word. Their approach, 
however, is not dynamical, and that is exactly the gap we want to fill out.

\bigskip

Our article is organized as follows: In 
Section~\ref{sec:channels} we introduce two different characterizations of 
\fmelo{completely positive and trace preserving (CPTP) linear maps},
which will allow us to describe generalized quantum dynamics and the 
coarse graining \fmelo{maps}. In this contribution, a coarse graining 
\fmelo{map} will simply be a \fmelo{CPTP linear map} that reduces the 
dimension of the system. Such maps were recently used to obtain a 
sufficient criteria for the entanglement of high-dimensional bipartite 
states~\cite{Ibrahim}. After that, in Section~\ref{sec:dynamics}, we obtain 
the effective dynamics $\Gamma_t$ induced by the coarse graining 
$\Lambda_\text{CG}$, underlying evolution $\g{U}_t$, and initial state 
$\psi_0$. The properties of the effective \fmelo{map} $\Gamma_t$ are 
discussed in Section~\ref{sec:properties}.  In Section~\ref{sec:distance} 
we show that the distance between two effective states may increase under 
the action of the same effective \fmelo{map} $\Gamma_t$. This is in 
contrast with the usual contractive property of \fmelo{CPTP linear 
maps}~\cite{nielsenchuang}. Finally, in Section~\ref{sec:conclusion} we 
draw some final conclusions and hint to some possible applications of the 
developed formalism.

\section{CPTP linear maps: general dynamics and coarse graining}
\label{sec:channels}

In order to define the coarse graining operations, which are the ones we are interested here, 
we will first briefly review some properties of CPTP linear maps. 
Comprehensive expositions can be found, for example, in~\cite{WatrousLect,MarkWildeBook,MichaelGuide,nielsenchuang}.

Let $\mc{H}_D\simeq \Cx^D$ be the Hilbert space assigned to a $D$-dimensional quantum system.  We define 
$\mc{L}(\mc{H}_D)$ as the set of all linear operators
acting on $\mc{H}_D$, and $\mc{D}(\mc{H}_D)= \{\psi \in \mc{L}(\mc{H}_D) 
|\; \psi\ge 0, \tr(\psi)=1\}$ the convex set  containing all the possible 
states of the system. \fmelo{Let $\Lambda:\mc{L}(\mc{H}_D) \rightarrow 
\mc{L}(\mc{H}_d)$ be} a linear map which  abides by two constraints: i) it 
is trace preserving, \fmelo{meaning that $\forall \psi \in 
\mc{L}(\mc{H}_D)$ we have $\tr(\psi)=\tr(\Lambda(\psi))$;} and ii) it is 
completely positive, i.e., \fmelo{for all positive operators $\psi \in 
\mc{L}(\mc{H}_D\otimes\mc{H}_Z)$, with $\mc{H}_Z$ an arbitrary finite 
dimensional Hilbert space,  the linear map 
$\Lambda\otimes\idty:\mc{L}(\mc{H}_D\otimes\mc{H}_Z) \rightarrow 
\mc{L}(\mc{H}_d\otimes\mc{H}_Z)$ is such that $\Lambda\otimes\idty(\psi)\ge 
0$} \cite{MichaelGuide}. The first 
imposition guarantees that probabilities are conserved through the 
\fmelo{map} action, while the completely positivity condition ensures that 
states are mapped into states even if the \fmelo{map} acts only on a 
subsystem of the whole system. The following well-known theorem 
gives a very  useful characterization of \fmelo{CPTP linear maps}.
\begin{theorem}[\cite{nielsenchuang,MichaelGuide}]
	A linear map $\Lambda:\mc{L}(\mc{H}_D) \rightarrow 
\mc{L}(\mc{H}_d)$ is \cris{completely positive and 
trace-preserving} if and only if 
there exists a finite 
set of linear operators $\{K_i\}_{i=1}^N$, with each $K_i:\mc{H}_D 
\rightarrow \mc{H}_d$ known as a Kraus operator, such that $\forall \psi \in \mc{L}(\mc{H}_D)$:
		\[\Lambda(\psi)= \sum_{i=1}^N K_i  \psi \\ K_i^\dagger \text{ with } \sum_{i=1}^N K_i^\dagger 
K_i = \mathds{1}_{D}.\]
\label{thm:kraus}
\end{theorem}
It is worth noticing that \fmelo{CPTP linear maps} generalize the evolution of a quantum system, with the unitary evolution being a particular \fmelo{linear map} $\g{U}_t:\mc{L}(\mc{H}_D) \rightarrow \mc{L}(\mc{H}_D)$ with a single Kraus operator, namely the unitary $U_t$ itself. In general, the number of Kraus operators is unlimited, but it is always possible to characterize a \fmelo{CPTP linear map} $\Lambda:\mc{L}(\mc{H}_D) \rightarrow \mc{L}(\mc{H}_d)$ with a set of Kraus operators with at most $D.d$ elements~\cite{nielsenchuang}, as this is the number of generators for the map. Moreover, the set of Kraus operators describing a given \fmelo{CPTP linear map} is not unique. Given the two sets $\{K_i\}_{i=1}^N$ and $\{K_i^\prime\}_{i=1}^M$, with $N\geq M$, they represent the same \fmelo{CPTP linear map} if, and only if, there exists a unitary $U \in \text{SU}(N)$ such that $K_i = \sum_j U_{ij}K_j^\prime$ (where, if necessary, we pad the 
smallest set with zeros)~\cite{nielsenchuang,MichaelGuide}.

This more general type of evolution allows for describing processes where there is a loss of information about the system, with pure states evolving to mixed ones. That is the case, for instance, when one is dealing with open quantum systems~\cite{breuer}. 

\begin{figure}[ht]
	\includegraphics[width=\linewidth]{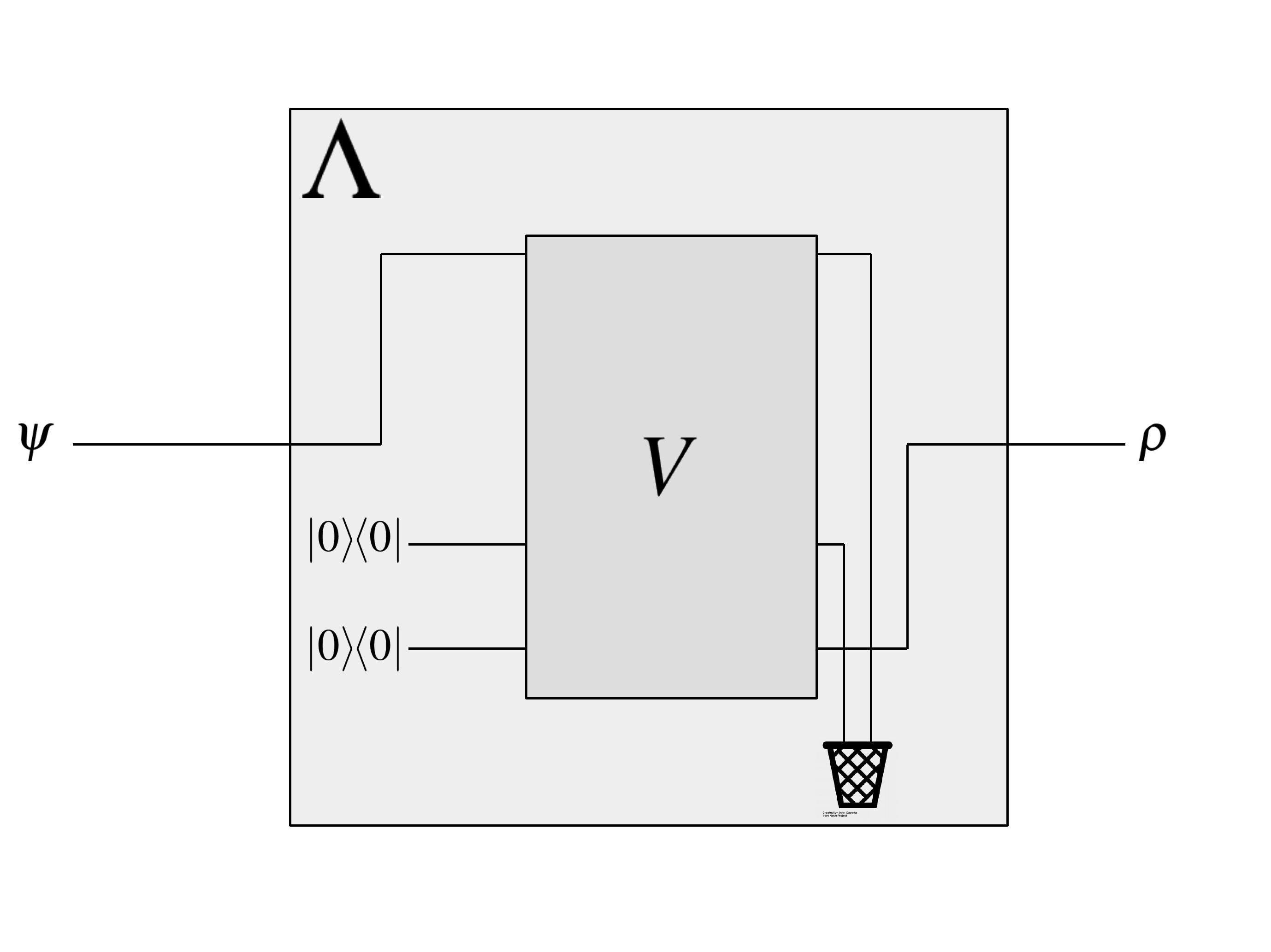}
	\caption{\footnotesize \textbf{Operational interpretation of a \fmelo{CPTP linear map} $\Lambda$.} }\label{fig:UnitaryStinespring}
\end{figure}

For the coarse graining operations we are going to employ below, the following (see Fig.~\ref{fig:UnitaryStinespring}) operational way to 
describe \fmelo{CPTP linear maps} will turn handy. 

\begin{theorem}[\cite{MichaelGuide}]
Let $\Lambda:\mc{L}(\mc{H}_D) \rightarrow \mc{L}(\mc{H}_d)$ be a \fmelo{CPTP linear map}. Then there exists an auxiliary Hilbert space $\mc{H}_r$, with dimension $r\leq d$, and a unitary $V$ acting on $\mc{H}_D\otimes \mc{H}_r \otimes \mc{H}_d$ such that $\forall \psi \in \mc{L}(\mc{H}_D)$ 
\[\Lambda(\psi)= \tr_{Dr}[V(\psi\otimes \proj{0} \otimes \proj{0})V^\dagger].\]
\label{thm:stinespring}
\end{theorem}
Operationally, this theorem means that we can interpret \fmelo{CPTP linear maps} $\Lambda$ as a unitary interaction among three systems, and further discarding of the first two  parties. See 
Fig.~\ref{fig:UnitaryStinespring}. This interpretation is  reminiscent of open quantum systems, where the system interacts unitarily with the environment, with the latter being discarded as we have no control about, or interest in, it. Here, however, the roles of system and environment are not so well delineated. As we want to allow for maps with different input-output dimensions, the partial trace is taken over the auxiliary system and also over the factor encoding the initial system state.

The theorems above provide equivalent characterization of \fmelo{CPTP linear maps}, hence we will use them interchangeably. In fact, it is easy to relate them by setting $\forall \ket{\psi} \in \mc{H}_D, \;V(\ket{\psi}\otimes\ket{0}\otimes\ket{0})=\sum_{i=1}^D \sum_{j=1}^r \ket{i}\otimes \ket{j}\otimes K_{ij}(\ket{\psi})$. This 
connection shows that the auxiliary system is necessary as to accommodate 
\fmelo{CPTP linear maps} which require a number of Kraus operators bigger than 
$D$. \cris{We should stress that} for a 
\fmelo{CPTP linear map} with a set of Kraus operators $\{K_i\}_{i=1}^N$, we take the dimension of $\mc{H}_r$ as 
$r=\ceil{N/D}$, and must find an equivalent set of Kraus operators with $D 
r$ elements, $\{K_i^\prime\}_{i=1}^{Dr}$.
\cris{Hence, whenever $N > D$ the auxiliary dimension $r$ will be greater 
than one}.

We are finally in position to establish the coarse graining operations. 
Roughly speaking, descriptions are named coarse-grained when some fine 
details of the underlying model are smoothed out, or replaced by average 
behaviors. In order to get valid descriptions of states after the coarse 
graining, we define it as a \fmelo{CPTP linear map} that reduces the dimension of the system: \[\Lambda_\text{CG}: 
\mc{L}(\mc{H}_D)\rightarrow\mc{L}(\mc{H}_d) \text{ with } D>d.\]
When one is not able to resolve the system in full detail, the coarse graining map gives an effective state for the system.

\begin{figure}[t]
	\includegraphics[width=\linewidth]{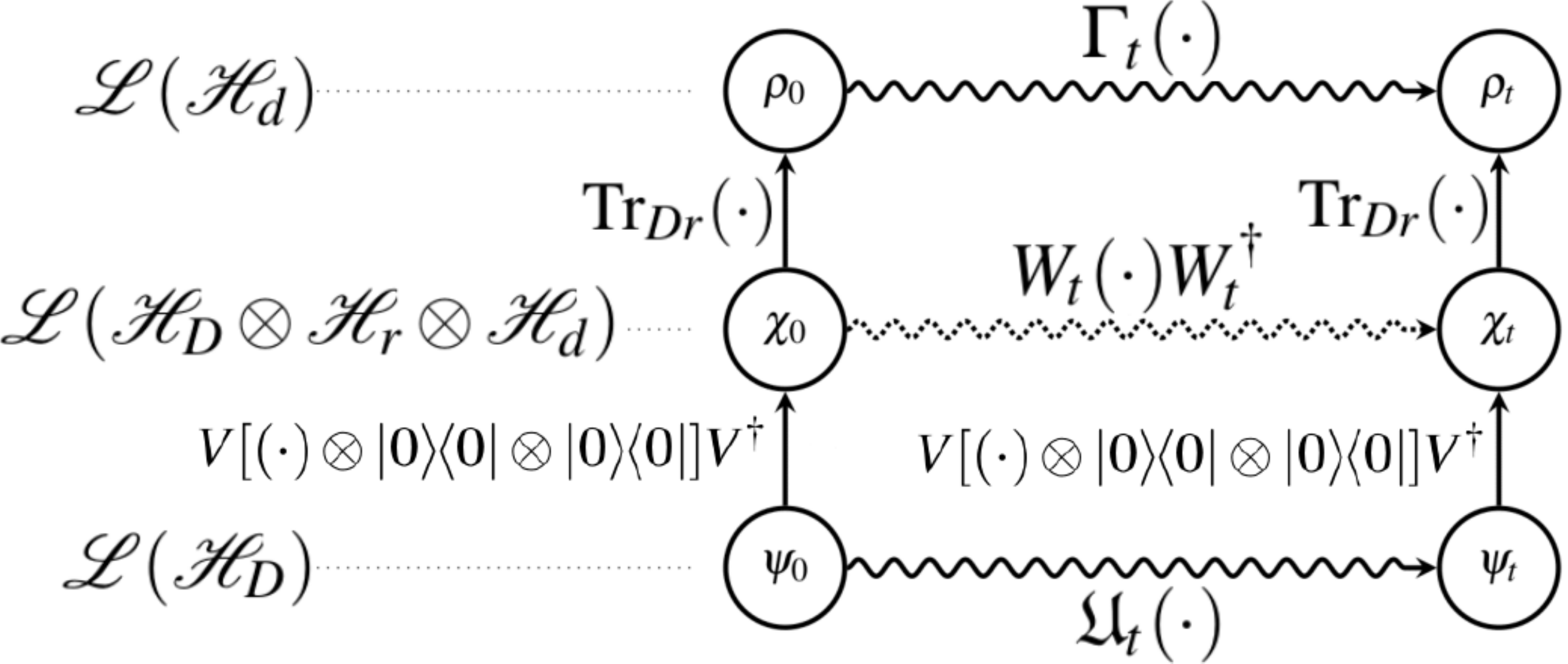}
	\caption{\footnotesize \textbf{The distinct levels and dynamics induced by the coarse graining  $\Lambda_\text{CG}$.}}\label{fig:CGlevels}
\end{figure}

Resorting to the characterization of \fmelo{CPTP linear maps} in Theorem~\ref{thm:stinespring}, we know that there exists an auxiliary space $\mc{H}_r$ and a unitary  $V:\mc{H}_D\otimes \mc{H}_r\otimes\mc{H}_d\rightarrow \mc{H}_D\otimes \mc{H}_r\otimes\mc{H}_d$, such that 
\[\Lambda_\text{CG}(\psi) = \tr_{Dr}[V(\psi\otimes \proj{0}\otimes\proj{0})V^\dagger].\]
Operationally, what the unitary $V$ accomplishes is to "write" the accessible degrees of freedom 
into the party in $\mc{H}_d$, while the unaccessible degrees of freedom are left in $\mc{H}_D\otimes \mc{H}_r$ to be later discarded. See 
Fig.~\ref{fig:CGlevels}. The intermediate states $\chi_0 = V(\psi_0\otimes \proj{0}\otimes\proj{0})V^\dagger$ and $\chi_t = V(\psi_t\otimes 
\proj{0}\otimes\proj{0}) V^\dagger$, which live in $\mc{H}_D\otimes \mc{H}_r\otimes\mc{H}_d$, are virtual states, in the sense that they are mathematical abstractions. In this level the two contributions of degrees of freedom, accessible and non-accessible are split, but may be correlated.

\subsection{Example: a blurred and saturated  detector}\label{subsec:detector}

In order to give a concrete example, let us consider a 
 typical optical lattice experiment~\cite{Kuhr10, Gross15, Greiner15}. In these 
experiments a periodic oscillating potential  is constructed by 
counter-propagating light beams, and individual atoms are trapped in each 
potential minimum. In the deep Mott insulator regime  two hyperfine levels of each atom act as a qubit,  and neighboring  qubits 
interact with each other via a Heisenberg-like Hamiltonian. The measurement 
of each atom is made via a fluorescence technique: the atoms are shone with 
a laser in way that if an atom is in the state, say, $\ket{1}$, light is scattered by the atom, whereas 
if its state is $\ket{0}$ no light is scattered. To resolve 
the light coming from each atom a powerful lens is necessary, and only 
recently a single-atom resolution was accomplished~\cite{KuhrReview}.

To simplify, consider the case with only two atoms. Suppose that the lens 
available is not good enough as to resolve the light coming from each 
individual atom. In this situation the states $\ket{01}$ and $\ket{10}$ 
cannot be distinguished. Moreover, imagine that the amount of light coming 
from a single atom is already sufficient to saturate the detector. Then, 
having two excitations, $\ket{11}$, or one excitation, $\ket{01}$ or 
$\ket{10}$, leads to the same signal. In such conditions to describe the 
experiment with two atoms is superfluous, and an effective description 
becomes handy. These experimental conditions suggest the coarse graining presented in Table~\ref{tb:CGtable}.

\begin{table}[ht]
\footnotesize
\begin{align*}
\Lambda_\text{CG}(\borb{00}{00})&=\proj{0} &\Lambda_\text{CG}(\borb{01}{00})&= \frac{\borb{1}{0}}{\sqrt{3}}\\
\Lambda_\text{CG}(\borb{00}{01})&= \frac{\borb{0}{1}}{\sqrt{3}}  &\Lambda_\text{CG}(\borb{01}{01})&= \borb{1}{1}\\
\Lambda_\text{CG}(\borb{00}{10})&= \frac{\borb{0}{1}}{\sqrt{3}} &\Lambda_\text{CG}(\borb{01}{10})&= 0\\
\Lambda_\text{CG}(\borb{00}{11})&=  \frac{\borb{0}{1}}{\sqrt{3}}  &\Lambda_\text{CG}(\borb{01}{11})&= 0\\ 
\\
\Lambda_\text{CG}(\borb{10}{00})&=  \frac{\borb{1}{0}}{\sqrt{3}} &\Lambda_\text{CG}(\borb{11}{00})&=  \frac{\borb{1}{0}}{\sqrt{3}} \\
\Lambda_\text{CG}(\borb{10}{01})&= 0 &\Lambda_\text{CG}(\borb{11}{01})&= 0\\
\Lambda_\text{CG}(\borb{10}{10})&= \borb{1}{1} &\Lambda_\text{CG}(\borb{11}{10})&= 0\\
\Lambda_\text{CG}(\borb{10}{11})&= 0 &\Lambda_\text{CG}(\borb{11}{11})&= \borb{1}{1}\\
\end{align*}
\normalsize
\caption{\label{tb:CGtable} \footnotesize \textbf{Coarse graining for a blurred and saturated detector.} If a detector does not distinguish between the two systems, and does not differ between one or two excitations, this coarse graining gives the effective description of the system.}
\end{table}

Note that as the detector does not distinguish between the states 
$\ket{01}$, $\ket{10}$, and $\ket{11}$ there can be no coherence in this 
subspace. Furthermore, the $1/\sqrt{3}$ factors are 
necessary to make $\Lambda_\text{CG}$ a \fmelo{CPTP linear map}. This signals that coherences in the 
effective description might decrease, but they do not necessarily 
vanish~\cite{FaistThesis}. This can be readily seen by evaluating the action of 
$\Lambda_\text{CG}$ over a general two-qubits pure state $\ket{\psi}= 
\sum_{i,j=0}^1 c_{ij}\ket{ij}$, with $c_{ij}\in \Cx$, which gives:
\footnotesize
\[\ \Lambda_\text{CG}(\proj{\psi}) = \begin{pmatrix}
|c_{00}|^2& c_{00}\frac{c_{01}^*+c_{10}^*+c_{11}^*}{\sqrt{3}}\\
c_{00}^*\frac{c_{01}+c_{10}+c_{11}}{\sqrt{3}}&|c_{01}|^2+|c_{10}|^2+|c_{11}|^2
\end{pmatrix}.\] 
\normalsize
This effective state accounts for the statistics of all possible measurements that can be carried out by the detector here modeled. It is thus the description that really matters for this experimental condition, not carrying unaccessible information.

The Kraus operators for this map can be easily obtained by a quantum process tomography~\cite{nielsenchuang}, and are given by:
\footnotesize
\begin{align*}
K_1&= \begin{pmatrix}
1&0&0&0\\
0&1/\sqrt{3}&1/\sqrt{3}&1/\sqrt{3}
\end{pmatrix};\\
K_2&= \begin{pmatrix}
0&0&0&0\\
0&1/\sqrt{3}&0&-1/\sqrt{3}
\end{pmatrix};\\
K_3&= \begin{pmatrix}
0&0&0&0\\
0&1/\sqrt{3}&-1/\sqrt{3}&0
\end{pmatrix};\\
K_4&= \begin{pmatrix}
0&0&0&0\\
0&0&1/\sqrt{3}&-1/\sqrt{3}
\end{pmatrix}.
\end{align*}
\normalsize
As we have four Kraus operators, $N=4$, and the dimension of the underlying system is also four, $D=4$, then the auxiliary system in $\mc{H}_r$ can be taken as 1-dimensional and as such can be ignored. With the above Kraus operators, and neglecting the system in $\mc{H}_r$, one can immediately obtain the corresponding unitary $V$ for this example:
\footnotesize
\begin{equation*}
V=\begin{pmatrix}
1&0&0&0&0&0&0&0\\
0&0&1/\sqrt{3}&0&1/\sqrt{3}&0&1/\sqrt{3}&0\\
0&1&0&0&0&0&0&0\\
0&0&1/\sqrt{3}&-1/\sqrt{3}&0&0&-1/\sqrt{3}&0\\
0&0&0&0&0&1&0&0\\
0&0&1/\sqrt{3}&1/\sqrt{3}&-1/\sqrt{3}&0&0&0\\
0&0&0&0&0&0&0&1\\
0&0&0&1/\sqrt{3}&1/\sqrt{3}&0&-1/\sqrt{3}&0\\
\end{pmatrix}.
\end{equation*}
\normalsize
 
\section{Coarse Graining  Induced Dynamics}\label{sec:dynamics}

Now we address the central question of this contribution: what are the dynamics that
might emerge from a fully quantum description if we are not able to resolve 
the system in all its details? More concretely,  we look for an effective 
map $\Gamma_t$ which makes the diagram in Fig.~\ref{FigureDiagramStates} 
consistent, i.e, in a way that $\rho_t\equiv\Gamma_t (\rho_0) = \Lambda_\text{CG} \circ 
\g{U}_t (\psi_0)$, with $\rho_0=\Lambda_\text{CG}(\psi_0)$. The induced 
dynamics then emerges from a coarse grained description of the underlying 
dynamics.

To obtain the induced dynamics $\Gamma_t$ acting on the effective state 
$\rho_0$, we generalize the procedure suggested by \v{S}telmachovi\v{c}  
and Bu\v{z}ek in~\cite{Buzek01}. There they proposed to write the state of the 
system and environment as the tensor product of its local parts plus a 
correlation term. Despite the fact that here we do not have such a splitting 
between system and environment, the action of the unitary $V$, see 
Fig.~\ref{fig:CGlevels}, suggests the following decomposition:
\beq
\chi_0=\left(\omega_0\otimes\rho_0\right) + \left(\chi_0 -\omega_0\otimes\rho_0\right),
\label{eq:decomposition}	
\eeq
where $\chi_0=V(\psi_0\otimes\proj{0}\otimes\proj{0})V^\dagger$, $\rho_0=\Lambda_\text{CG}(\psi_0)=\tr_{Dr}(\chi_0)$, $\omega_0=\tr_d(\chi_0)$. Note that $\omega_0$ is a state in $\mc{D}(\mc{H}_D\otimes\mc{H}_r)$. Equation~\eqref{eq:decomposition} is equivalent to \v{S}telmachovi\v{c}  and Bu\v{z}ek decomposition in the abstract level $\mc{H}_D\otimes\mc{H}_r\otimes\mc{H}_d$, with the last term now representing the correlation between the degrees of freedom which can be accessed and those that  cannot. 
As $V$ is unitary, we can equivalently write:
\begin{align}
\psi_0\otimes\proj{0}\otimes\proj{0} = &V^\dagger\left(\omega_0\otimes\rho_0\right)V +\nonumber\\
&+ V^\dagger\left(\chi_0 -\omega_0\otimes\rho_0\right)V.
\label{eq:Psi0decomposition}
\end{align}

From the left hand side of Eq.~\eqref{eq:Psi0decomposition} we get the evolved effective state by applying the underlying evolution map $\g{U}_t$ onto the first tensor factor, followed by the application of $V$ and further partial trace of the two first tensor factors:
\begin{align*}
\rho_t & = \Lambda_\text{CG}\circ \g{U}_t (\psi_0)\\
  &= \tr_{Dr}\big[V\big(\g{U}_t(\psi_0)\otimes\proj{0}\otimes\proj{0}\big)V^\dagger\big]\\
  &= \Gamma_t(\rho_0).
\end{align*}
The last equality comes from demanding consistence of the diagram in Fig.~\ref{FigureDiagramStates}. Accordingly, assuming the underlying evolution map  of the form $\g{U}_t(\cdot)=U_t(\cdot)U_t^\dagger$, from the right hand side of Eq.~\eqref{eq:decomposition} we get the effective evolution:
\begin{align}
\Gamma_t(\rho_0)= &  
\tr_{Dr}\big(W_t\left(\omega_0\otimes\rho_0\right)W_t^\dagger\big)+ 
\nonumber\\
&+\tr_{Dr}\big(W_t\left(\chi_0 
-\omega_0\otimes\rho_0\right)W_t^\dagger\big), 
\label{eq: AboveEvolutionEquation}
\end{align}
where $W_t = V.(U_t\otimes \idty\otimes \idty).V^\dagger$ is
the unitary  evolution operator in the level 
$\mc{H}_D\otimes\mc{H}_r\otimes\mc{H}_d$, i.e., $\chi_t = W_t \chi_0 
W_t^\dagger$. See Fig~ \ref{fig:CGlevels}. 

The above evolution equation can be rewritten in a more meaningful way as
\beq
\Gamma_t(\rho_0)= \sum_{i,j}M_{ij}\rho_0M_{ij}^\dagger + \tr_{Dr}\big(W_t\left(\chi_0 -\omega_0\otimes\rho_0\right)W_t^\dagger\big),
\label{eq:evol}
\eeq
with $M_{ij}=\sqrt{p_j}(\bra{\phi_i}\otimes\idty)W_t(\ket{\phi_j}\otimes\idty)$, where we employed the spectral decomposition  $\omega_0 = \sum_j p_j \proj{\phi_j}$. This is the dynamics that emerges if one is not able, or does not wish, to resolve all the details of the underlying system.

The expression in Eq.~\eqref{eq:evol} is composed by two 
contributions: the first one displays a Kraus form (see 
Theorem~\ref{thm:kraus}), with $\{M_{ij}\}$ the corresponding set of 
effective Kraus operators; the second one represents the evolution of the 
correlations between accessible and non-accessible degrees of freedom.  This 
second term can be more clearly appreciated by evoking the Bloch 
representation of $\chi_0$:
\begin{align}
\chi_0= \frac{1}{Drd}&\big(\idtym_{Dr}\otimes\idtym_{d}+\idtym_{Dr}\otimes \vec{\alpha}.\vec{\sigma}_d+\big.\\&\big.+ \vec{\beta}.\vec{\sigma}_{Dr} \otimes \idtym_{d} + \sum_{i,j} \theta_{ij}\sigma_{Dr}^{(i)}\otimes \sigma_{d}^{(j)}\big),\nonumber
\end{align}
where  
$\vec{\sigma}_q=(\sigma_q^{(1)},\sigma_q^{(2)},\ldots,\sigma_q^{(q^2-1)})^T$ 
is a vector whose components are the $q \times q$ generalized Pauli 
matrices,  $\vec{\alpha}\in \Rl^{d^2-1}$ is the Bloch vector of $\rho_0$,  
$\vec{\beta}\in \Rl^{(Dr)^2-1}$ is the Bloch vector of $\omega_0$, and the 
$((Dr)^2-1)(d^2-1)$ coefficients $\theta_{ij}\in \Rl$ fix the correlation between accessible and 
non-accessible degrees of freedom. Defining the correlation matrix 
$[\Theta]_{ij}=(\theta_{ij}-\beta_i \alpha_j)/Drd$, the evolution of the 
coarse grained state can be written as:
\beq
\Gamma_t(\rho_0)= \sum_{i,j}M_{ij}\rho_0M_{ij}^\dagger + \sum_{i,j}\Theta_{ij}\tr_{Dr}\big( W_t\sigma_{Dr}^{(i)}\otimes \sigma_{d}^{(j)}W_t^\dagger\big).
\label{eq:evol_correlation}
\eeq
It can be easily verified that $\sum_{i,j}M_{ij}^\dagger 
M_{ij}=\mathds{1}_{d}$, and that $\tr_d\left(\tr_{Dr}\big( 
W_t\sigma_{Dr}^{(i)}\otimes \sigma_{d}^{(j)}W_t^\dagger\big)\right)=0$ as 
$W_t$ is unitary and the (generalized) Pauli matrices are traceless. These conditions
guarantee that $\tr_d(\Gamma_t(\rho_0))=1$ for all  times. The structure 
of this type of evolution is very similar to the one describing open quantum 
systems when system and environment are initially correlated~\cite{Carteret08, 
Buzek01}.

\subsection{Example: effective dynamics as seen by a blurred and saturated detector}\label{subsec:eff_dynamics}

Consider again the situation described in  subsection~\ref{subsec:detector}: 
two atoms in neighboring wells of 
an optical lattice being observed by a blurred and saturated detector. 
Suppose now that the atoms interact as specified by the Hamiltonian 
$H= \hbar J \sigma_{z}\otimes\sigma_{z}$, with $J$ a coupling constant in units of frequency. In such situation, an initial two-qubit pure state 
$\ket{\psi_0}= \sum_{i,j=0}^1 c_{ij}\ket{ij}$ evolves to:
\[\ket{\psi_t}= (c_{00}\ket{00}+c_{11}\ket{11})\text{e}^{-iJt}+(c_{01}\ket{01}+c_{10}\ket{10})\text{e}^{iJt}.\]

\begin{figure}
	\includegraphics[width=\linewidth]{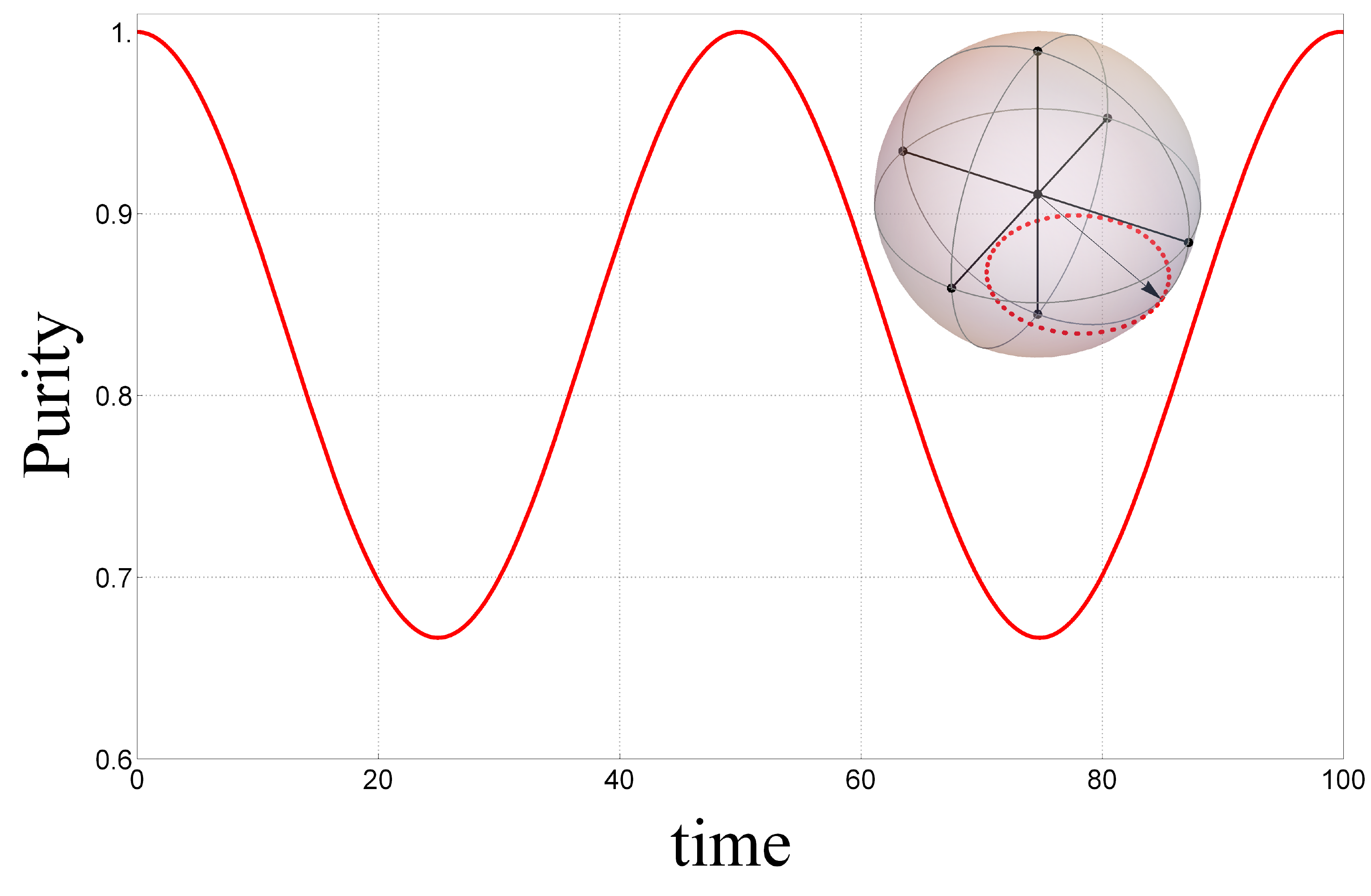}
	\caption{\footnotesize \textbf{Effective evolution as seen by a blurred-saturated detector.} The plot shows an oscillatory behavior for the purity of the effective state. The inset shows the effective state trajectory in the Bloch sphere.}\label{fig:purity}
\end{figure}

The evolution of the effective state can then be easily evaluated via $\rho_t=\Lambda_\text{CG}(\psi_t)$, to give:
\footnotesize
\[\ \rho_t = \begin{pmatrix}
|c_{00}|^2& c_{00}\frac{\text{e}^{-2\ii J t}(c_{01}^*+c_{10}^*)+c_{11}^*}{\sqrt{3}}\\
c_{00}^*\frac{\text{e}^{2\ii J t}(c_{01}+c_{10})+c_{11}}{\sqrt{3}}&|c_{01}|^2+|c_{10}|^2+|c_{11}|^2
\end{pmatrix}.\] 
\normalsize
As a concrete example, the effective evolution of a state $\psi_0$ with all coefficients $c_{ij}$ equal, $c_{ij}=1/2$ for $i,j\in\{0,1\}$, is shown in the inset 
of Fig.~\ref{fig:purity}. Figure \ref{fig:purity}  also shows how the purity, 
$\tr(\rho_t^2)$, oscillates with time, exhibiting the alternation between 
pure and mixed state in the effective level. This is in clear contrast with 
the complete description of the system, where the system is pure for all times.

It is interesting to notice that the coefficients of $\psi_0$  define the state $\rho_0$, but also enter in the definition of the effective \fmelo{map} $\Gamma_t$. In the above example this can be verified by evaluating $\omega_0=\tr_{Dr}\chi_0$, which will also depend on the coefficients $c_{ij}$. That in turn, means that  the effective Kraus operators $M_{ij}$ will also change with the $c_{ij}$ -- thus by changing $\rho_0$ the \fmelo{map} may change. The same is true for the  correlation  matrix $\Theta_{ij}$. This interdependence of the parameters is treated in the next section, where the properties of $\Gamma_t$ are analyzed.

\section{Properties of $\Gamma_t$}\label{sec:properties}
The effective \fmelo{map} $\Gamma_t$ is generated by the underlying evolution 
$\g{U}_t$, the coarse graining map $\Lambda_\text{CG}$, and the state $\psi_0$. Equation~\eqref{eq:evol_correlation}, however, 
does not make explicit how the \fmelo{map} depends on the elemental state 
$\psi_0$. For instance, how do we change the effective \fmelo{map} $\Gamma_t$ 
for a fixed input state $\rho_0$? Or, how to change the effective input state 
keeping $\Gamma_t$ fixed? In what follows we address these and other questions making use of  the Bloch representation 
for $\psi_0$:
\beq
\psi_0=\frac{1}{D}\big(\idtym_D +\vec{\gamma_0}.\vec{\sigma_D}\big),
\eeq
where $\vec{\gamma_0}\in \Rl^{D^2-1}$ is the Bloch vector of $\psi_0$.

\subsection{Fix $\rho_0$, change $\Gamma_t$}

Fixed the coarse graining map, the Bloch vector $\vec{\alpha}$ of $\rho_0$ is obtained  from $\vec{\gamma_0}$ by the linear relations:
\beq
\left\{\begin{array}{ccc}
\alpha_1 &=& \tr[\Lambda_\text{CG}(\psi_0 (\vec{\gamma_0}))\sigma_{d}^{(1)}];\\
\alpha_2 &=& \tr[\Lambda_\text{CG}(\psi_0 (\vec{\gamma_0}))\sigma_{d}^{(2)}];\\
\vdots&\vdots&\vdots\\
\alpha_{d^2-2} &=& \tr[\Lambda_\text{CG}(\psi_0 (\vec{\gamma_0}))\sigma_{d}^{(d^2-2)}];\\
\alpha_{d^2-1} &=& \tr[\Lambda_\text{CG}(\psi_0 (\vec{\gamma_0}))\sigma_{d}^{(d^2-1)}].\\
\end{array}\right.
\label{eq:hyperplanes}
\eeq
In the $D^2-1$ dimensional space of Bloch vectors $\vec{\gamma_0}$ of 
$\psi_0$, these constraints represent hyperplanes whose intersection depicts 
the effective state $\rho_0$. It is important to notice that since $D>d$, 
the set of linear equations for the coefficients $\alpha_j$ is 
under-determined, meaning that various states $\psi_0$ lead to the same 
effective state $\rho_0$. Geometrically, in the ``$\gamma$-space'', 
this many-to-one mapping is visualized as an hyper-surface of possible 
solutions.

\begin{figure}
	\begin{tabular}{c}
	\includegraphics[width=0.8\linewidth,valign=t]{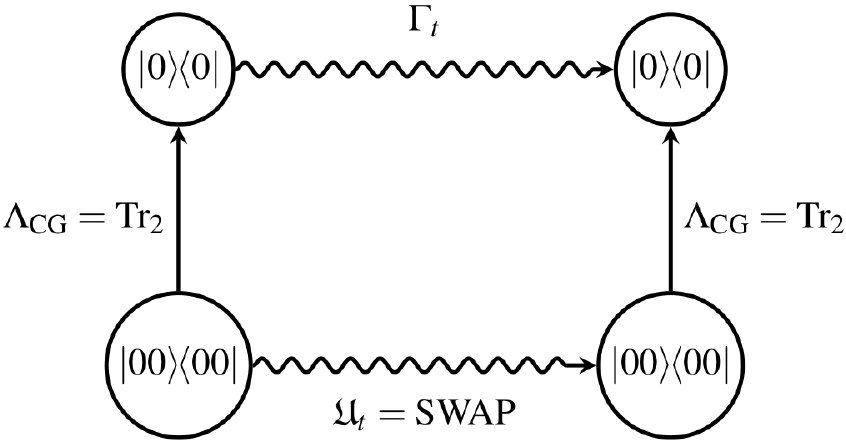}\\
	\\
	\\
	\includegraphics[width=0.8\linewidth,valign=t]{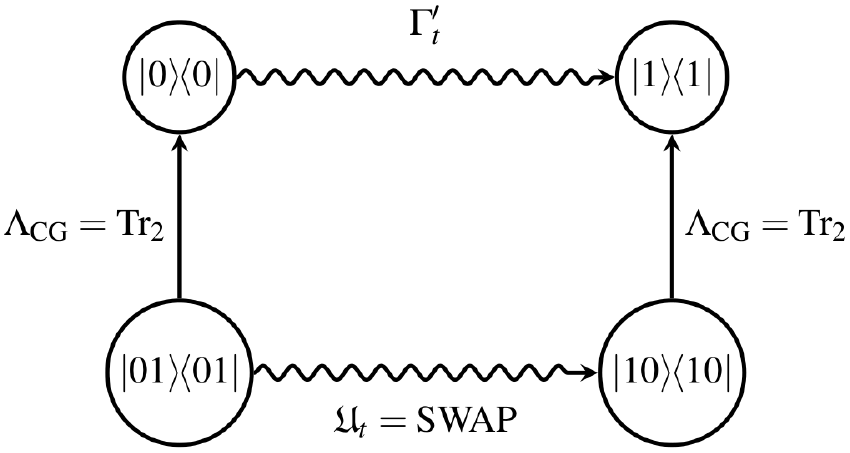}	
	\end{tabular}
	\caption{ \footnotesize \textbf{Simple example of fixing $\rho_0$ and changing the effective \fmelo{map}.} For the case where we fix the unitary mapping as the $SWAP$, i.e., $U_t\ket{ij}=SWAP\ket{ij}=\ket{ji}$, and the coarse graining as the usual partial trace on the second component, we see that different underlying states generate different effective \fmelo{maps}. The fact that the emergent \fmelo{maps} cannot be the same is clear as if that was the case the same input would lead to two different outputs. }\label{fig:swap}
\end{figure}

\begin{figure}
	\begin{tabular}{l}
	\includegraphics[width=0.6\linewidth]{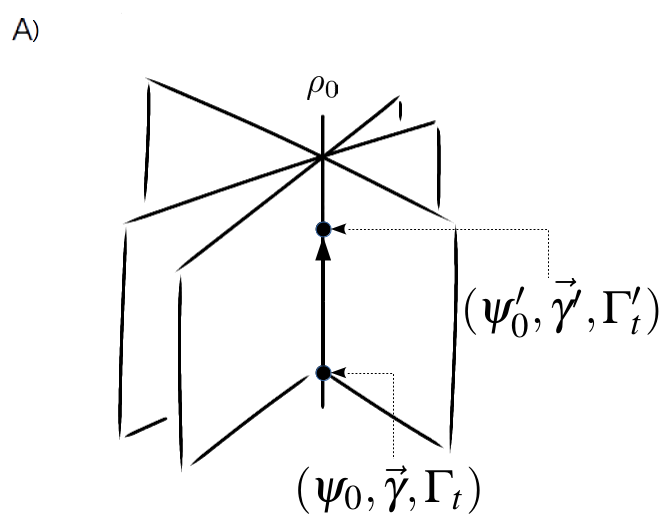}\\
	\includegraphics[width=\linewidth]{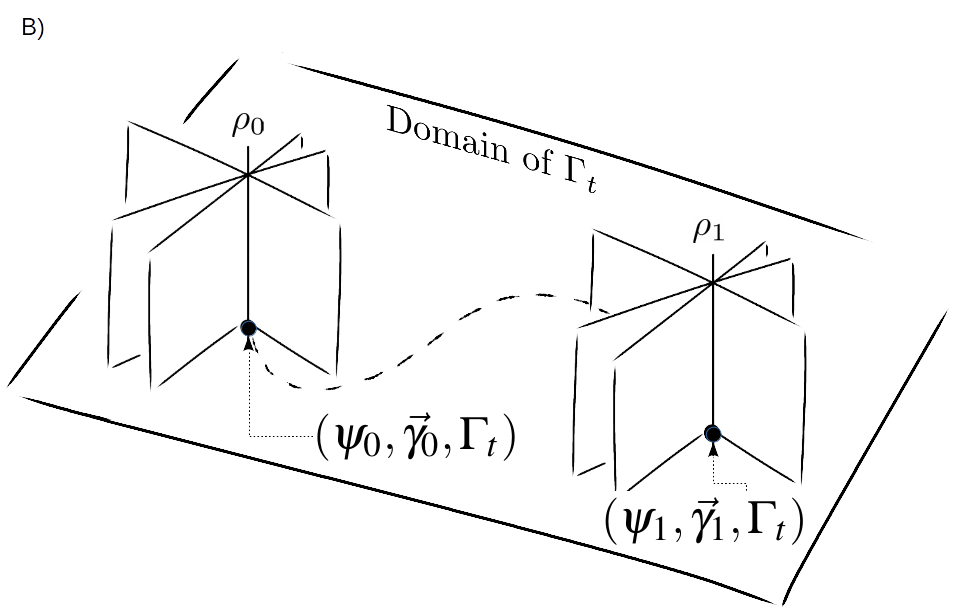}
	\end{tabular}
	\caption{ \footnotesize \textbf{The $\gamma$-space, and the effects of changing the underlying state.} The intersection between the hyperplanes defines the effective state. A non point-like region reflects the fact that many underlying states lead to the same effective state. A) The top panel shows a change in the underlying state that does not change the effective state $\rho_0$. This change may, nevertheless, have impact on the effective \fmelo{map}.  B) Changing the underlying state such that its Bloch vector $\vec{\gamma}$ moves normally to the hyperplanes, changes only the effective state preserving the effective \fmelo{map}. The region obtained by such normal displacement of the Bloch vector defines the domain of an effective \fmelo{map}. }\label{fig:GammaSpaceMoves}
\end{figure}

Now, with this geometric perspective in mind, a fixed coarse 
graining and a fixed underlying evolution, it can be seem 
that changes in $\psi_0$, that move $\vec{\gamma_0}$ parallel to the 
hyper-planes within the solution hyper-surface  will not affect the  
effective state $\rho_0$. Nevertheless, such change can 
induce modifications in $\omega_0$  or in $\Theta$, and as such 
$\Gamma_t$ will change. A simple example is presented in 
Fig.~\ref{fig:swap}, and an abstract representation of the $\gamma$-space 
and the change in $\Gamma_t$ can be seen in Fig.~\ref{fig:GammaSpaceMoves}.

\subsection{Fix $\Gamma_t$, change $\rho_0$: the domain of $\Gamma_t$}\label{subsec:same_channel}

To change the effective state $\rho_0$ we must change the value of the  
$\alpha_i$'s. Geometrically this is represented by moving the hyperplanes, defined in Eq.~\eqref{eq:hyperplanes}, in the $\gamma$-space. After the hyper-planes 
displacement, a new intersection is obtained representing now another 
effective state, say, $\rho_1$. As moving $\vec{\gamma_0}$ parallel to the 
hyper-planes changes the \fmelo{map}, this time we must move $\vec{\gamma_0}$ 
perpendicular to the hyper-planes. This guarantees that only the effective 
state is changing. See Fig.~\ref{fig:GammaSpaceMoves}A).

It is important to notice that this change might in fact modify $\omega_0$  
or $\Theta$. This, however, comes only because of the change in the 
effective input state, as these quantities might be functions of 
$\vec{\alpha}$. Putting it differently, the effective Kraus operators might change, but this is only due to the change in the input of the effective map $\Gamma_t$. The dynamical equation~\eqref{eq:evol_correlation} can be rewritten as to make this dependence explicit:
\beq
\Gamma_t(\rho_0(\alpha))= \sum_{i,j}M_{ij}(\alpha)\rho_0(\alpha) {M_{ij}^{\dagger}}(\alpha) + \zeta(\alpha),
\label{eq:evol_alpha}
\eeq
where $\zeta(\alpha)=\sum_{i,j}\Theta_{ij}(\alpha)\tr_{Dr}\big( 
W_t\sigma_{Dr}^{(i)}\otimes \sigma_{d}^{(j)}W_t^\dagger\big)$. 

We can now  determine the domain of a given $\Gamma_t$. An 
effective \fmelo{map} $\Gamma_t$ is generated by an 
elemental state $\psi_0$, an underlying evolution 
map $\g{U}_t$, and a coarse graining \fmelo{map} $\Lambda_\text{CG}$. This 
information already gives the first element in the domain of $\Gamma_t$, 
namely, $\rho_0=\Lambda_\text{CG}(\psi_0)$. The coarse graining 
$\Lambda_\text{CG}$ fixes the hyper-planes in the $\gamma$-space through 
Eq.\eqref{eq:hyperplanes}. Let $\vec{v}_i$ be the normal vector for the 
$i$-th hyper-plane, and $\vec{\gamma_0}$ be the Bloch vector of $\psi_0$. 
The domain of $\Gamma_t$ is then given by all $\rho=\Lambda_\text{CG}(\psi)$ 
generated from $\psi$  with Bloch vector $\vec{\gamma}$ for which there exists 
coefficients $c_i\in \Rl$ such that $\vec{\gamma} =\vec{\gamma_0} +\sum_i 
c_i \vec{v}_i$. The latter condition guarantees that the Bloch vector of all states in the domain of $\Gamma_t$ can be reached from $\vec{\gamma_0}$ by moving it perpendicular to the hyper-planes in Eq.~\eqref{eq:hyperplanes}, and as such not changing the effective \fmelo{map}.

This immediately implies that the domain of $\Gamma_t$ is convex: Let 
$\rho_a=\Lambda_\text{CG}(\psi_a)$ and $\rho_b=\Lambda_\text{CG}(\psi_b)$ be 
in the domain of $\Gamma_t$. This means that there exists coefficients 
$ \{a_i\} \subset \Rl $  and $ \{b_i\} \subset \Rl$ such that 
the Bloch vectors of $\psi_a$ and $\psi_b$ can be written as 
$\vec{\gamma}_a=\vec{\gamma_0} +\sum_i a_i \vec{v}_i$ and 
$\vec{\gamma}_b=\vec{\gamma_0} +\sum_i b_i \vec{v}_i$, respectively. There 
are many states in $\mc{L}(\mc{H}_D)$ which after the coarse graining lead 
to the convex combination $\rho=p \rho_a + (1-p) \rho_b$, with $p\in[0,1]$. 
In particular, the state $\psi= p \psi_a + (1-p) \psi_b$ is such that 
$\Lambda_{CG}(\psi)=\rho$ and it has Bloch vector $\vec{\gamma_0} +\sum_i 
(pa_i+(1-p)b_i) \vec{v}_i$. Therefore the convex combination $\rho$ is also 
in the domain of $\Gamma_t$.

\subsection{Effective positivity and complete-positivity of $\Gamma_t$}

Equation~\eqref{eq:evol_alpha} clearly shows that in general 
$\Gamma_t$ is not of ``Kraus'' form, like shown in Theorem~\ref{thm:kraus}. 
This means that if $\Gamma_t$ is taken as a map between states from $\mc{D}(\mc{H}_d)$ to itself, then $\Gamma_t$ 
is not completely positive, possibly not even positive. 

However, as we have just seen, the domain of a given $\Gamma_t$ is not necessarily all the states in $\mc{D}(\mc{H}_d)$. Restricting the action of $\Gamma_t$ to its domain guarantees the positivity of the \fmelo{map}. This can be immediately verified by the simple consistence of the diagram in Fig.~\ref{fig:CGlevels}, which demands 
$$\Gamma_t(\rho_0)=\Gamma_t\circ \Lambda_\text{CG}(\psi_0)=\Lambda_\text{CG}\circ\g{U}_t(\psi_0).$$
As the right most part of this equation is a composition of 
positive \fmelo{maps}, then the positivity of the first term is also 
guaranteed. 

The same line of thought can be used as to argument for the complete 
positivity of $\Gamma_t$. Indeed, as we are constructing our framework upon 
quantum mechanics, no contradiction with it can be obtained.   However this 
argument should  not go through without a caveat: not all extensions of 
effective states $\rho$ in the domain of a given $\Gamma_t$ into $\omega\in 
\mc{D}(\mc{H}_A\otimes\mc{H}_d)$, with $\mc{H}_A$ the Hilbert space of an 
auxiliary system, are possible. The possible extensions for 
$\rho=\Lambda_\text{CG}(\psi)$ are those that can be obtained from states 
$\Psi\in\mc{D}(\mc{H}_A\otimes\mc{H}_D)$ such that $\tr_A(\Psi)=\psi$, which 
guarantees that $\tr_A(\idty_A\otimes\Lambda_\text{CG}(\Psi))=\rho$, and 
that $\psi$ generates the \fmelo{map} $\Gamma_t$ (together with the underlying 
evolution $\g{U}_t$). We call such a family of states $\Psi$ as the \textit{set of effective complete positivity} for 
$\Gamma_{t}$. Physically this constraint comes from the fact that 
if one does not have control of all the degrees of freedom of a system, 
then not all the states can be generated. In another perspective, the 
entanglement that can be created in the fundamental level 
$\mc{H}_A\otimes\mc{H}_D$ is, in general, decreased by the action of the 
coarse graining map~\cite{Ibrahim}.

\begin{figure}
	\includegraphics[scale=0.4]{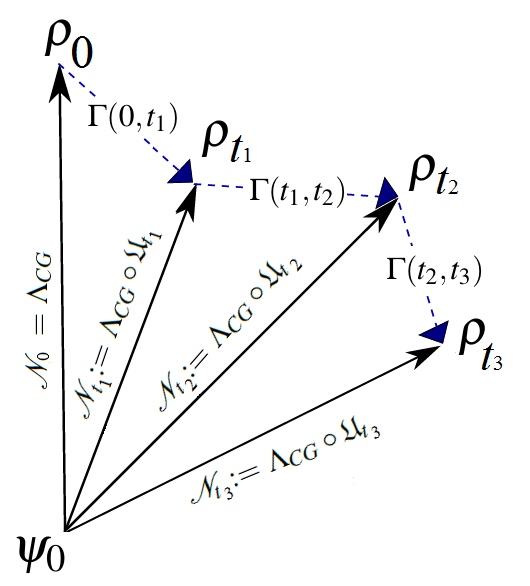}
	\caption{\footnotesize \textbf{Recovering strict complete positivity for the effective \fmelo{map}.} One way to obtain a family of CP effective \fmelo{maps} is to require the \fmelo{map} $\mc{N}_{t_k}=\Lambda_{CG}\circ \g{U}_{t_k}$ to be CP-divisible.}\label{fig:CPdivisible}
\end{figure}

Strict complete positivity can be re-obtained if we demand that the composite \fmelo{map} 
$\mc{N}_t:=\Lambda_{CG}\circ\g{U}_t$ to be CP-divisible~\cite{buscemi}. If  that is the case, the definition of CP-divisible \fmelo{maps} requires $\mc{N}_{t_{k}} = 
\Gamma_{(t_{k},t_{j})}\circ\mc{N}_{t_{j}}$ for all $t_k\ge t_j$, with 
$\Gamma_{(t_{k},t_{j})}$, the effective evolution \fmelo{map} for the time 
interval $[t_j,t_k]$, completely positive \cris{(see 
Fig.~\ref{fig:CPdivisible})}. This shows a connection between the theory 
of coarse grained maps and the theory of non-Markovian maps~\cite{buscemi, 
RivasHuelgaPlenio}.

\section{Consequence: Effective distance increase by $\Gamma_t$}\label{sec:distance}

A common property of \fmelo{CPTP linear maps}~\eqref{thm:kraus}, is that the distance between two input states cannot increase. Mathematically, let 
$\Lambda:\mc{L}(\mc{H}_D)\rightarrow \mc{L}(\mc{H}_d)$ be a \fmelo{CPTP linear map} and $\psi$ and $\psi^\prime$ be states in $\mc{L}(\mc{H}_D)$. Then 
$||\Lambda(\psi)-\Lambda(\psi^\prime)||_1\le||\psi-\psi^\prime||_1$, where the 1-norm is defined as $||A||_1:=\tr(\sqrt{A^\dagger A})$. Physically, this means, for instance, that the discrimination between two unknown quantum states cannot be improved by any further processing of the states~\cite{nielsenchuang}.

The effective \fmelo{map} $\Gamma_t$, as discussed in the previous section, is not in general of Kraus form. Can then the distance between two effective states increase? As argued before, no contradiction with quantum mechanics can arise. In fact, it is simple to check that the distance between two effective states is upper-bounded, for all times, by the distance between the underlying initial states. Let $\rho_0=\Lambda_{CG}(\psi_0)$ and $\rho_0^\prime=\Lambda_{CG}(\psi_0^\prime)$ be effective states in $\mc{L}(\mc{H}_d)$ with respective evolved states $\rho_t=\Gamma_t(\rho_0)$ 
and $\rho_t^\prime=\Gamma_t(\rho_0^\prime)$. Then
\begin{align}
||\rho_t-\rho_t^\prime||_1&=||\Lambda_{CG}(\psi_t)-\Lambda_{CG}(\psi_t^\prime)||_1;\nonumber\\
&\le||\psi_t-\psi_t^\prime||_1;\label{eq:distance_t}\\
&=||\g{U}_t(\psi_0)-\g{U}_t(\psi_0^\prime)||_1;\nonumber\\
&\le||\psi_0-\psi_0^\prime||_1.\label{eq:distance_0}
\end{align}
The last inequality turns into an equality in the case of a unitary mapping $\g{U}_t$, i.e, $\g{U}_{t}(.)=U_{t}(.)U^{\dagger}_{t}$ for some unitary $U_t$.

\begin{figure}
	A)\includegraphics[width=0.8\linewidth,valign=t]{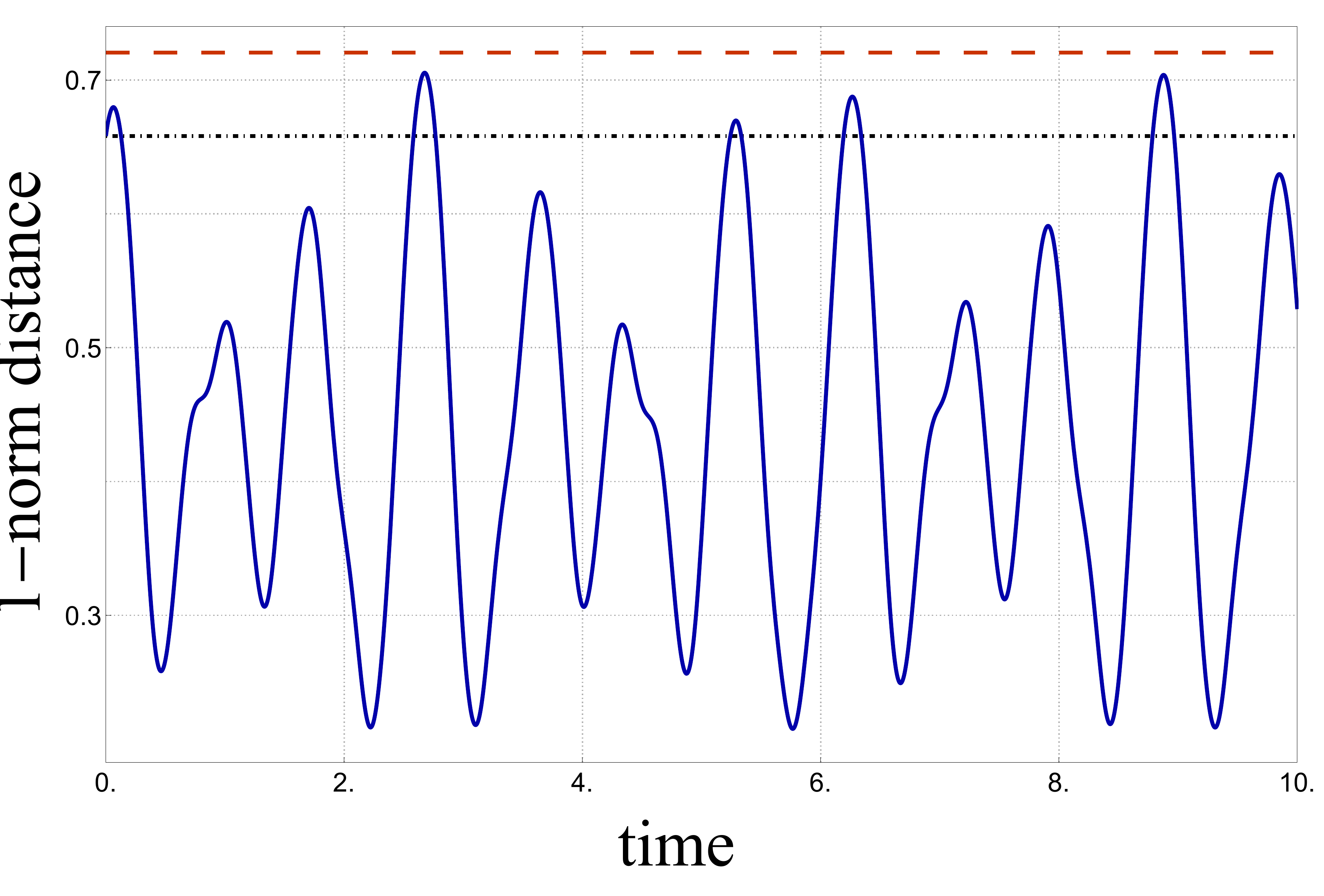}\\
	B)\includegraphics[width=0.8\linewidth,valign=t]{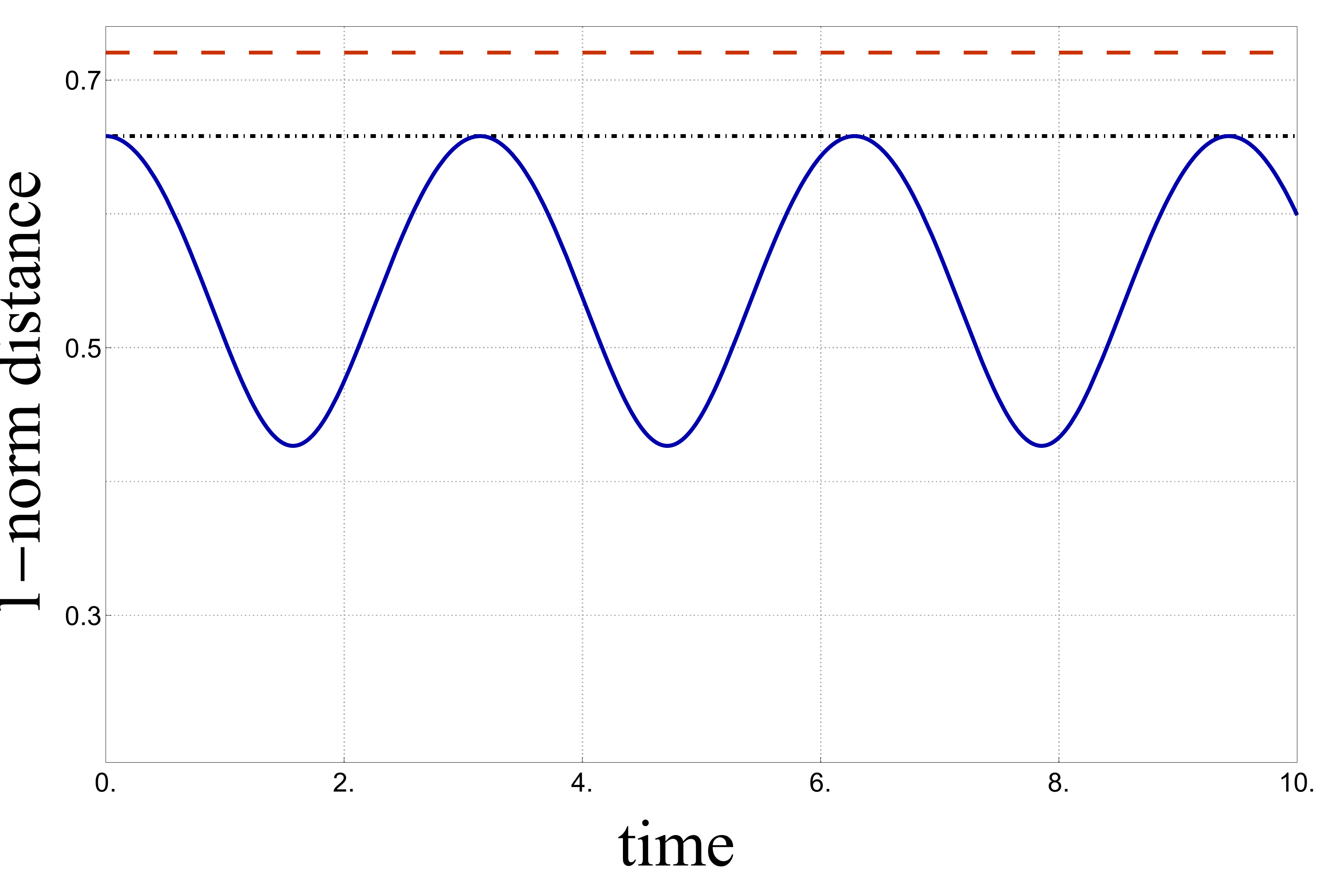}
	\caption{\footnotesize \textbf{Distance increase for the effective dynamics.} In the above plots $||\rho_t-\rho^\prime_t||_1$, $||\psi_0-\psi^\prime_0||_1$, and  $||\rho_0-\rho^\prime_0||_1$ are represented, respectively by the blue-continuous line, red-dashed line, and the black-dot-dashed line. Contrary to the usual contractive property of \fmelo{CPTP linear maps}, on the effective level the distance between two states undergoing the same process may increase. This increase is, however, upper-bounded by the distance between the underlying states (red-dashed line). A) The underlying interaction is dictated by the Hamiltonian $H=\hbar J\sigma_{z}\otimes\sigma_z$. We see that the distance oscillates, increasing for some time intervals. Nevertheless, in this case, we always have $||\rho_0-\rho_0^\prime||_1 \ge ||\rho_t-\rho_t^\prime||_1$. B) The underlying evolution is dictated by the Hamiltonian  $H=\hbar J\sigma_{z}\otimes\sigma_z+3(\sigma_{x}\otimes\idtym+ \idtym\otimes\sigma_x)$. In this case we see that  $||\rho_t-\rho_t^\prime||_1$ can even go beyond $||\rho_0-\rho_0^\prime||_1$.}
	\label{fig:distEvol}
\end{figure}

This, however, does not imply that a distance increasing between effective states is not allowed. In fact, it is possible to have an increase in distance between the effective states undergoing the same effective \fmelo{map}. Take for example the coarse graining  describing the  blurred-saturated detector (\ref{subsec:detector}), the underlying dynamics given by the Hamiltonian $H=\hbar J\sigma_{z}\otimes\sigma_z$ (\ref{subsec:eff_dynamics}), and select two states $\psi_0$ and $\psi_0^\prime$ which generate the same effective \fmelo{map}~\ref{subsec:same_channel}. Figure~\ref{fig:distEvol}A) shows the distance evolution between the  two effective states $\rho_t= \Lambda_\text{CG}(\psi_t)$ and $\rho_t^\prime= \Lambda_\text{CG}(\psi_t^\prime)$. A clear oscillation of the distance is observed. In this example, nevertheless, we have that $||\rho_0-\rho_0^\prime||_1 \ge ||\rho_t-\rho_t^\prime||_1$ for all times. 

Now, switch on a transversal field, turning the Hamiltonian into $H=\hbar J\sigma_{z}\otimes\sigma_z+3(\sigma_{x}\otimes\idtym+ \idtym\otimes\sigma_x)$, and take initial states $\psi_0$, and $\psi_0^\prime$ in a way to have the same effective \fmelo{map}. The evolution of the distance between the effective states is shown in Fig.~\ref{fig:distEvol}B). In this case we see that the distance $||\rho_t-\rho_t^\prime||_1$ can go beyond $||\rho_0-\rho_0^\prime||_1$ for some specific times.

\section{Conclusion}\label{sec:conclusion}

When dealing with complex many-body quantum systems, the full description 
of the system and of its dynamics is prohibitive. Even in principle, it 
assumes that one has access to all the system's exponentially (in the 
number of constituents) many degrees of freedom. A ``simple" system 
composed of $60$ qubits, would require in general the measurement of about 
$(2^{60})^2\approxeq 1.33 \times 10^{36}$ observables to be fully 
characterized -- even if each measurement is performed in one femtosecond, 
this would take more than 3000 times the age of the universe to be 
accomplished. This is only for the state, the characterization of the 
dynamics is far more complex. Effective descriptions are thus mandatory in 
order to perceive macroscopic systems.

Pursuing the direction of effective descriptions,
here we investigated what types of dynamics may emerge from a 
full quantum description when one does not have access to, or is not interested in, all 
the degrees of freedom of a given system.  The presented formalism 
generalizes the theory  of open quantum systems, as it works also for 
closed systems. Here the split between system and environment is 
substituted by the split between accessible and non-accessible degrees of 
freedom. The possibility of correlations between these two types of degrees 
of freedom may generate effective dynamics that are not of Kraus form -- 
without violating any principle of quantum mechanics. This, in  turn, allowed for the distance 
between two effective states to increase under the action of a fixed 
effective \fmelo{map} -- in contrast to what is achievable in the 
underlying quantum description. 

Other aspects of this effective dynamics can be further explored. Most notably, the fact that the "Kraus operators" may depend on the state the \fmelo{map} is acting on. This suggests a possible way to explain how  non-linear dynamics may emerge from the quantum linear description: if one looks at the system only at time intervals for which the term quantifying correlations between accessible and non-accessible degrees of freedom, $\zeta(\alpha)$ in Eq.~\eqref{eq:evol_alpha}, vanishes, i.e., for a coarse grained time~\cite{Nadja17}, then the non-linearity of the first term may become apparent. This (possible) non-linearity, together with the distance increase between effective states undergoing the same effective \fmelo{map} may be the key to explain how chaotic systems arise from the underlying quantum mechanical description. 

Lastly, we hope that the formalism here presented can shed some light on 
the quantum-to-classical transition: the higher the  "zoom out" (stronger 
coarse graining, in the sense of larger difference between $D$ and $d$), 
the more simplified becomes the description of the system and its 
dynamics, 
with quantum features fading away. We believe that these ideas can be of 
interest for areas as quantum thermodynamics -- which tries to explain the 
thermalization of closed quantum systems~\cite{Eisert15, FaistThesis} --, 
and even to address the measurement problem in quantum 
mechanics~\cite{caslav15}.



\begin{acknowledgments}
	Special thanks goes to Fred Brito, for countless discussions, and careful reading of the manuscript.  We also thank R.~C.~Drumond, D.~Jonathan, and R.~L.~de Matos Filho for stimulating discussions. This work is supported by the Brazilian funding agencies  CNPq and CAPES, and it is part of the Brazilian National Institute for Quantum Information.
\end{acknowledgments}

%

\end{document}